*Disruptive effects of chlorpyrifos on predator-prey interactions of Ceratophrys ornata tadpoles: consequences at the population level using computational modeling.*

Salgado Costa Carolina[1,3], Rimoldi Federico[1,3], Pantucci Saralegui Morena Johana[1,3,4], Rubio Puzzo María Leticia[5], Trudeau Vance Lionel[6], Natale Guillermo Sebastián[2,3]

[3]Centro de Investigaciones del Medio Ambiente (CIM), CONICET-UNLP, Departamento de Química, Facultad de Ciencias Exactas, Universidad Nacional de La Plata, Bv. 120 nº 1489 (1900), La Plata, Buenos Aires, Argentina.
CCT CONICET La Plata, Consejo Nacional de Investigaciones Científicas y Técnicas, B1904CMC La Plata, Argentina. E-mail: csalgadocosta@quimica.unlp.edu.ar, frimoldi@quimica.unlp.edu.ar, morena.pantucci@ilpla.edu.ar; gnatale@quimica.unlp.edu.ar

[4] Instituto de Limnología Dr. Raúl A. Ringuelet (ILPLA), CONICET- UNLP, Boulevard 120 y 62 (1900), La Plata, Buenos Aires, Argentina.

[5]Instituto de Física de Líquidos y Sistemas Biológicos (IFLYSIB), CONICET-UNLP, Departamento de Física, Facultad de Ciencias Exactas, Universidad Nacional de La Plata, 1900 La Plata, Argentina. Calle 59 no. 789 (1900), La Plata, Argentina.
CCT CONICET La Plata, Consejo Nacional de Investigaciones Científicas y Técnicas, B1904CMC La Plata, Argentina. E-mail: lrubio@iflysib.unlp.edu.ar

[6]Department of Biology, University of Ottawa, Ottawa, ON K1N 6N5, Canada. E-mail: trudeauv@uottawa.ca

1 Equal contribution
2 Corresponding author

**Abstract**

Large-scale ecotoxicological studies have technical and ethical limitations, both related to the need to expose large numbers of individuals to potentially harmful compounds. The computational modeling is a complementary useful and predictive tool that overcomes these limitations. Considering the increasing interest in the effects of pesticides on behavioral traits, the aim of this study was to evaluate the effects of chlorpyrifos (CPF) on intra- and inter-specific interactions of anuran tadpoles, complementing traditional ecotoxicological tools with a theoretical analysis verified by computational simulations. Experiments were developed under two consecutive phases: a first phase of exposure (treated and control group), and a second phase of interactions. The second phase consisted of evaluating the effects of CPF on intra- and inter-specific interactions of exposed *C. ornata* (*Co*) tadpoles acting as predators and unexposed *Rhinella fernandezae* (*Rf*) tadpoles acting as prey (Experiment I), under different predator-prey proportions (0/10 = 0*Co*-10*Rf*, 2/8, 4/6, 6/4, 8/2, 10/0). Also, intraspecific interactions of three *Co* tadpoles under different conditions of exposure were evaluated (Experiment II: 3 exposed *Co*, 2 exposed *Co*/1 non-exposed, 1 exposed *Co*/2 non-exposed). During the exposure phase, chlorpyrifos induced significant mortality from 48 h (48h: $p < 0.05$, 72h-96h: $p < 0.001$), irregular swimming, tail flexure, and the presence of subcutaneous air. Also, it induced effects on the sounds emitted after 96 h of exposure, registering less number of pulses and higher dominant frequencies, and altered intra- and inter-specific interactions. During the interaction phase, the larvae continued to show sound effects, however, the antipredator mechanism continued to be operating and efficient. Finally, it was possible to model the behavior of the larvae under the effects of chlorpyrifos. We conclude that experimental data and computational modeling matched within errors. Therefore, computational simulation is a valuable ecotoxicological tool that provides new information and allows predicting natural processes.

**Main findings**: Environmentally-relevant chlorpyrifos concentrations disrupt predator-prey interactions of *Ceratophrys ornata* tadpoles.

# 1. Introduction

Ecotoxicology addresses the study of the effects of environmentally relevant pollutants on ecosystems. Currently, this discipline has numerous validated tools and procedures that allow assessments of the chemical toxicity, including emerging pollutants, and evaluating a wide range of effects from the gene to higher levels of ecological organization (e.g., population, community) (Newman 2014). Although pesticides were one of the first classes of pollutants to be assessed globally, there is still a lack of knowledge about their effects in some groups of animals, especially at the population level and beyond. In the field of herpetological ecotoxicology, the effects of the regularly used and moderately toxic organophosphate insecticide chlorpyrifos have been documented. However, these studies mainly reported frequently used endpoints at the genotoxic, biochemical, and individual level (Cowman and Mazanti 2000, Richards and Kendall 2002, El-Merhibi, Kumar, and Smeaton 2004, Kerby 2006, Yin et al. 2009, Bernabò et al. 2011, Silva et al. 2020). Particularly in Argentina, only 5 native species out of 169 (= 3%) have been evaluated (Ruiz de Arcaute et al. 2012, Sotomayor et al. 2012, Attademo et al. 2015, Liendro et al. 2015, Sotomayor et al. 2015, Sansiñena et al. 2018, Quiroga et al. 2019, Barreto et al. 2020), and only three of those references reported low environmentally-relevant concentrations disrupting key behaviors which may affect the fitness of a particular species. Here, we focused our attention on *Ceratophrys ornata*, an Argentinian horned frog. We previously discovered for the first time in any vertebrate that: (1) the tadpoles of *C. ornata* are able to emit sounds during premetamorphic stages (Natale et al., 2011); (2) sounds are emitted during conspecific interactions as part of an antipredator mechanism that diminishes the frequency of predation between conspecifics in the presence of heterospecific prey; (3) tadpoles never emit sounds during interspecific interactions when acting as predators (Natale et al. 2011, Salgado Costa 2016); (4) sounds could play an important role in population dynamics and survival, and (4) chlorpyrifos disrupts sounds (sound duration, number of pulses and dominant frequency) emitted by *C. ornata* tadpoles; therefore, effects on sounds are a very good sublethal endpoint and a promising possible biomarker (Salgado Costa et al. 2018).

Chlorpyrifos is a broad-spectrum organophosphate insecticide widely used to control agricultural pests (Testai, Buratti, and Di Consiglio 2010). The partition coefficient (log $K_{ow}$ ~ 5) suggests a high potential to bioaccumulate. It enters aquatic ecosystems by drift and runoff occurring in water, suspended particles and sediment (Barron and Woodburn 1995, Jergentz et al. 2005, Marino and Ronco 2005, Claver et al. 2006, Mac Loughlin, Peluso, and Marino 2017, Alvarez et al. 2019). Thus, aquatic species are among the most vulnerable biota.

Particularly amphibians have several characteristics that make them particularly sensitive to pollutants (such as pesticides), and thus represent good bioindicators or sentinels of local conditions (DeGarady and Halbrook 2006, Sansiñena et al. 2018). Considering that only a few studies have evaluated behavioral effects of predators and prey all together (Relyea and Edwards 2010), we now can assess sound production during predator-prey interactions in our unique model system. Adequate experimental design using the traditional ecotoxicological tools has technical and ethical limitations: it requires a detailed analysis of many simultaneous interactions (tens to hundreds), a non-disturbing video recording system coupled with a hardware with adequate data storage, and the exposure of many living organisms that will likely perish during the course of LC50 testing. The advance of knowledge and the possibility of developing new techniques in accordance with the care and welfare of laboratory animals, call for novel and rigorous ecotoxicological procedures. In this context, computational simulations are a useful tool to understand complex systems that cannot be experimentally modeled, constituting a complement to both theoretical and experimental studies. In the last half-century, computational modeling has extended the application range to solve problems from Exact and Natural Sciences to Social Sciences and/or Economy (Landau and Binder 2014, Winsberg 2019). Models are simplified representations of reality, elaborated on a manageable scale with the objective of describing the response of more complex and therefore realistic systems. The application of these models as predictive tools of natural processes is an alternative to the development of large-scale ecotoxicological studies.

Contemplating both the lack of knowledge and the increasing interest in the effects of pesticides on behavioral traits, the aim of this study was to evaluate the effects of chlorpyrifos on intra- and inter-specific interactions of *C. ornata* and other anuran tadpoles, using both the traditional ecotoxicological tools and theoretical models analyzed by means of computational simulations. In this context, we aimed to test the theoretical model and evaluate its usefulness. This tool could serve as a basis for decision-making in pest management under the paradigm of a more empathic experimentation and a more sustainable agriculture.

**2. Materials and methods**

**2.1 Study species: breeding and maintenance**

Tadpoles of *Ceratophrys ornata* (Anura: Ceratophryidae) have been selected as a promising ecotoxicological model. The species is distributed in the Pampas region of Argentina and Southern Brazil, and inhabits grasslands, agroecosystems and urban areas (Cei 1980, Gallardo 1974). It was selected since it shares certain conspicuous characteristics with some species of the genus: a) tadpoles emit audible sounds that have been described as part of an antipredator mechanism (Natale et al. 2011, Salgado Costa et al. 2014, Salgado Costa 2016); b) the frequency of intra- and inter-specific interactions strongly depends on the availability of prey (Salgado Costa et al. 2014, Salgado Costa et al. 2016, Salgado Costa 2016); c) the conservation status of the species is "vulnerable"(Natale and Salgado Costa 2012); and d) tadpoles are highly sensitive to chlorpyrifos which induced both lethal and sublethal effects (morphological abnormalities, growth inhibition, swimming and sound alterations) (Salgado Costa et al. 2018).

Adults of *C. ornata* were collected in the field (Collection permit: 22500-14357/11, Decree 209/11 and 14/12). Tadpoles were obtained by inducing spawning with the Amphiplex method (Trudeau et al. 2010). Eggs and tadpoles were reared under controlled laboratory conditions (Salgado Costa et al. 2018). Before the start of the experiments, until tadpoles reached stage 31 (Gosner, 1960), they were fed *ad libitum* with naturally encounter food items (tadpoles of different native species, and pieces of fish), and beef liver, as a dietary supplement in case of lack of native prey, all according to Salgado Costa (2016) protocol. All procedures for the care and use of laboratory animals were performed in agreement with local guidelines for vertebrate animal welfare (Protocol Number 023-22-15).

**2.2 Experimental stage**

The design consisted of an experimental stage to obtain information followed by a theoretical stage (see next section). During the experimental stage two independent experiments were run under two consecutive phases: a first phase of exposure, with a treated group (tadpoles under the effects of chlorpyrifos) and a control group; and a second phase of intra- and inter-specific interactions (interaction phase). Experiment I consisted of evaluating intraspecific and interspecific interactions of previously exposed *C. ornata* tadpoles under different predator-prey proportions; Experiment II consisted of evaluating in detail intraspecific interactions of *C. ornata* tadpoles placed together and previously exposed under different conditions. The methodology and controlled laboratory conditions followed pre-published procedures (Salgado Costa et al. 2018).

**2.2.1 Exposure phase**

The toxicity of the insecticide chlorpyrifos was assessed using the active ingredient (95.1% of purity) provided and certified by Gleba S.A. (CAS number 2921-88-2). Previous studies showed that nominal and measured concentrations (expected and observed values, respectively) did not differ (Salgado Costa et al. 2018, Barreto et al. 2020).

All tests were performed following international recommendations to accomplish with test procedures (ASTM, 2002; USEPA, 2002), with minor modifications for *C. ornata* (Salgado Costa et al., 2018). During the exposure phase, tadpoles from stage 31 (Gosner 1960) were exposed individually during 96 h in 200 ml glass jars to 0 (control group) and 0.02 mg/L of chlorpyrifos (treated group). The concentration 0.02 mg/L is the LOEC for swimming alterations and effects on sounds (Salgado Costa et al. 2018). Mortality, swimming alterations, presence of morphological abnormalities, and prey consumption were recorded every 24 h according to Salgado Costa et al. (2016, 2018). Briefly, individuals were considered dead when immobility, rapid decomposition of the body and/or absence of cardiac activity were corroborated; swimming alterations were classified according to the descriptions made by Brunelli et al. (2009), and morphological abnormalities according to the categories proposed by Bantle et al. (1996). Prey consumption was determined by registering the proportion of individuals who ate the supplied prey (tadpoles of native species) per treatment. Sounds were recorded and analyzed following validated procedures previously established by Salgado Costa et al. (2018). Briefly, each larva was recorded individually and at random for 30 s. Then, sounds emitted at 5, 10 and 20 s of recording were selected and sound duration (Sd) in seconds (s), number of pulses (Np) and dominant frequency (Df) in Hertz (Hz) were determined.

**2.2.2 Interaction phase: Experiment I and II**

During the interaction phase, experiments were carried out using dechlorinated water without insecticide. The aim of Experiment I was to assess the effects of chlorpyrifos under different predator-prey proportions. Considering that cannibalism is higher at a higher proportion of predators (=lower proportion of prey) at a fixed density (Salgado Costa et al. 2014), we aimed to determine the existence of the antipredator mechanism under the effects of the insecticide. The experiment was carried out during 48 h following Salgado Costa et al. (2014) with minor modifications: it consisted of a control group (n = 100) with *C. ornata* tadpoles individually exposed to 0 mg/L of chlorpyrifos, and a treated group (n = 100) with tadpoles individually exposed to 0.02 mg/L of chlorpyrifos during the exposure phase. An independent factor (= predator-prey proportion) with six levels and three replicates per level was assessed. Each treatment combined larvae of *C. ornata* (*Co*, predator) with those of *Rhinella fernandezae* (*Rf*, prey), all at the same density (4.35 tadpoles/L). The selected species is a natural prey of the larvae of *C. ornata*. Prey was not exposed to chlorpyrifos. Tadpoles were placed in plastic trays of 29x25x6 cm, totaling 18 experimental units. The following predator-prey proportions were evaluated: 0/10 (= 0*Co*-10*Rf*), 2/8, 4/6, 6/4, 8/2, 10/0. The number of consumed prey (= interspecific interactions) and the occurrence of cannibalistic events (conspecific interactions) at a particular time were recorded every 15 min during the first two hours, every 30 min during the following 4 h and every 2 h until the end of the experiment at 48 h. These recordings permitted the calculation of the following variables: 1) time to eat the first prey, 2) time to eat the second prey, 3) time to eat all prey, 4) time without eating (after depletion of all prey), 5) time to eat a congener (cannibalism event), 6) number of prey eaten, and 7) number of events of cannibalism. Finally, individuals of *C. ornata* from different experimental units were randomly selected and recorded at 0 h of interaction (= 96 h of exposure) and 24 h of interaction. Sounds were recorded and analyzed following validated procedures as previously mentioned.

For Experiment II, we simultaneously observed and audio-monitored intraspecific interactions of *C. ornata* tadpoles. The aim was to evaluate in detail the effects of chlorpyrifos on those interactions, and the efficiency of the antipredator mechanism. Tadpoles were observed and recorded during time lapses of 15 min (total observation time = 7 h 30 min). Three conditions with 10 replicates (= 30 experimental units) were assessed: a) interaction of 3 individuals of the control group (unexposed to chlorpyrifos during the exposure phase); b) interaction of 1 individual of the control group with 2 individuals of the treated group (exposed to chlorpyrifos during the exposure phase); and c) interaction of 2 individuals of the control group with 1 individual of the treated group. Prior to the running of the experiments, individuals belonging to each group were randomly selected, and their most conspicuous differences in size and color were recorded to identify them during monitoring. In all cases, individuals that attacked were recorded as predators, and victims as prey. Considering that each experimental unit consisted of 3 interacting individuals, it was possible to visually monitor them without any inconvenience. The resulting emission/non-emission of sounds, and the number of events of cannibalism and predator-predator encounters (= intraspecific) were quantified.

**2.3 Theoretical stage: Modeling and simulation**

The aim of the theoretical stage was to bring an additional tool to study the effects of chlorpyrifos on intra- and inter-specific interactions of *C. ornata* tadpoles. First, some simple models -explained in detail below- based on experimental observations were proposed. Secondly, these models were tested by Computational Simulations, and finally, the modeling results were compared with those of the experiments performed. The study is based on the Monte Carlo method (MC). The underlying concept of MC simulations is to understand the behavior of a given system of particles characterized by unknown parameters, many of which are difficult to determine experimentally. Results are computed based on repeated random sampling, and statistical analysis allow solving deterministic problems (for more details see e.g., Landau and Binder 2014). In this sense, the models are based on the particle-concept by focusing on tadpole's interactions with their neighbors.

Based on observations made during the experimental stage, and previously published results (Salgado Costa et al. 2014, Salgado Costa et al. 2016, Salgado Costa 2016), *C. ornata* larval movements in a confined environment are random, and intra- and inter-specific interactions depends on the availability of prey. This means that the frequency of predation between conspecifics significantly diminishes when prey items of another species are available. In this context, we aimed to elucidate the theoretical model which best describes intra- and inter-specific interactions of *C. ornata* (*Co*) tadpoles and *R. fernandezae* tadpoles (*Rf*, natural prey). If such a model is elucidated, then we would be able to corroborate our results from theoretical predictions and make other predictions on larger scales. We started the computational analysis with a simple and well-known model: The Random Walk Model on a lattice. Particles are initially placed randomly on a square lattice L x L and have the same probability to jump to their nearest neighbors if the site is empty. Open boundary conditions were applied to simulate the confinement of tadpoles during the experimental stage. Monte Carlo simulations started by randomly placing a given number of *Co*-like-particles ($N_{Co}$ predators) and *Rf*-like-particles ($N_{Rf}$, prey) on a square lattice of size L=100 and keeping constant $N=N_{Co}+N_{Rf}=10$. To allow comparisons with data from the experimental stage, we also fixed the predator-prey proportions: $N_{Co}/N_{Rf}$ = 2/8, 4/6, 6/4, 8/2, 10/0. Note that we previously tested several system sizes (L) in the range L=[50 , 200] but results were equivalent by changing the time scale, so we fixed L=100 without losing information. Regarding time scale, it is worth mentioning that a time-simulation unit in MC studies, the so-called "Monte Carlo step" (MCs), is defined when all N particles of the sample are, on average, once selected.

We focused on *Co-Co* and *Co-Rf* interactions evaluating the existence of an antipredator mechanism (APM). Therefore, we ran three Random Walk Models: one without APM (Model I), other with the presence of an APM (Model II), and the third complements the second one (Model III = Model II + insecticide), simulating the behavior of the exposed individuals. The interaction rules of models I and II can be summarized as follows: 1) at t = 0, $N_{Co}$ and $N_{Rf}$ particles are randomly placed on a square lattice and their probability of moving from one site to another is the same; 2) at time t > 0 each particle at (x,y) jumps with probability 1/4 to a near-neighbor site $(x',y')=(x\pm 1, y\pm 1)$ if the site is empty; 3) if the site $(x',y')$ is occupied, then three scenarios are possible: 3.1) if $Co \rightarrow Rf$ or $Rf \rightarrow Co$ occurs, the predator eats the prey with probability = 1, occupying the prey' site, 3.2) If $Rf \rightarrow Rf$ occurs, then each particle stays at each site, 3.3) When $Co \rightarrow Co$ occurs, the difference between the two models is implemented. Under Model I (without

APM), one of the two *Co*-particles is removed, with equal probability, and the other site remains occupied independently of the presence of prey. However, under Model II (with APM), only if $N_{Rf}$ (t) = 0 (absence of prey) then one of the *Co*-particle randomly selected is removed, being its site occupied by the other *Co*-particle. This means that cannibalism occur only if prey is unavailable as previously reported (Salgado Costa et al. 2014, Salgado Costa 2016).

The last model proposed, called Model III, adds to Model II (with APM) the sublethal effect of chlorpyrifos, simulating in a simple way the irregular swimming of tadpoles observed during the exposure phase, by introducing the possibility to stay in a site instead of moving. In this case, the randomly selected tadpole can either jump to an empty neighbor or remain in a site with a 1/5 probability.

Finally, we determined the number of particles (prey and predators) as a function of time (in MCs), the time needed to eat all prey as a function of the initial number of predators, for the different $N_{Co}/N_{Rf}$ and models considered, and the number of predator-predator encounters (=intraspecific interactions).

## 2.4 Statistical analysis

The level of significance set was 0.05 for all tests. Homogeneity of variances and normality were corroborated with Bartlett's and Shapiro-Wilk's test, respectively.

Results of the exposure phase were compared by two independent samples comparing tests (control versus treated group using Student t test or Mann-Whitney U test, as appropriate) considering lethal and sublethal variables (mortality, swimming activity, morphological abnormalities, and prey consumption).

Results of the interaction phase (Experiment I) were compared within groups (control and treated groups) and between groups (control versus treated groups) considering all variables (time to eat the first/second/all prey, time without eating, time to eat a congener and number of events of cannibalism) and predator-prey proportions (2/8, 4/6, 6/4, 8/2, 10/0). Comparisons within groups were made by univariate parametric (ANOVA, F) or non-parametric tests (Kruskal-Wallis, H), as appropriate. The comparison of those variables at a time between treated and control group was carried out by a Wilcoxon test for paired data.

Comparisons between bioacoustic variables (Sd, Np, Df) of the sounds emitted by larvae of the control group and the treated group were also performed by independent samples comparing tests (Student t test or Mann-Whitney, as applicable). Finally, a 2 x 2 contingency table between the categories (emission / non-emission of sounds) and groups (control / treated) was used to analyze intraspecific interactions of Experiment II.

The accuracy of MC simulation was estimated according to standard method to determine the error in a measure (Landau and Binder 2014). As it is well known, in the MC method the statistical error of a given magnitude f can be estimated by measuring the variance of *f*, defined as $\mathrm{Var}(f) \equiv \sqrt{\langle f^2 \rangle - \langle f \rangle^2}$, and it decreases with the square root of the number of independent samples $N_{samp}$. In our case, and from the direct analysis of the variance of the number of consumed particles (prey and predators), we estimated the number of independent trials needed to have an error less than the 5% of the measurement, being $N_{samp}$=100 enough to assure the reliability of the results (Supplementary Figure 1). Finally, results from the simulations and the experimental stage were compared by independent samples comparing tests (using Student t test or Mann-Whitney U test, as appropriate), paired t tests or by a chi-square test ($X^2$).

## 3. Results and discussion

### 3.1 Experimental stage: Exposure phase

Results are summarized in Table 1. Chlorpyrifos induced highly significant cumulative mortality compared to the control group from 48 h of exposure (48h: U = 26.00, p < 0.05; 72h and

96h: U = 0.00, p < 0.001). When considering species sensitivity distribution, a recent study published an acute risk analysis and indicated that shrimps, cladocerans, and amphipods species are the most sensitive taxa to chlorpyrifos, whereas other arthropods, but also mollusks and vertebrates are affected at higher concentrations (Alvarez et al. 2019). In that study, the estimated acute HC5 was 0.064 µg/L, which is three orders of magnitude lower than the concentration tested here to induced lethal and sublethal effects (= 0.02 mg/L). However, when considering only the anuran taxa and their LC50-96h data available for chlorpyrifos, the species sensitivity distribution previously published by Salgado Costa et al. (2018) shows that *C. ornata* is a sensitive species close to the 29th percentile.

At the low concentration tested, chlorpyrifos also induced swimming alterations, which were noted as irregular swimming, defined as erratic swimming, body twisting and convulsions according to descriptions made by Brunelli et al. (2009). This altered activity was registered in all the exposed organisms at the end of the exposure phase. This may be related with the neurotoxic action of chlorpyrifos exerted by inactivation of acetylcholinesterase, which causes behavioral alterations in aquatic vertebrates (Brewer et al. 2001, Kavitha and Rao 2008). In addition, chlorpyrifos induced morphological abnormalities like tail flexure (a bending generated from the bottom or in the middle region of the tail by moving it side), and the presence of subcutaneous air (the abnormal presence of one or more air bubbles in the visceral cavity, on either or both sides of the body) as described by Sansiñena et al. (2018). The presence of this abnormality could be related to the reported irregular swimming. Also, a significant reduction of prey consumption at 96 h was detected (U = 15.942, p < 0.0001).

Finally, the insecticide also induced significant effects on the sounds emitted after 96 h of exposure (Table 1), registering less number of pulses (U = 179.5, p < 0.0005, 3 less pulses, on average) and a higher dominant frequency (U = 4.00, p < 0.0005, 2000 Hz higher on average) in tadpoles exposed to chlorpyrifos; no significant differences were found for sound duration (t = 1.01, df = 62, p = 0.318). Although sound alterations as a result of chlorpyrifos exposure have been previously reported in this species (Salgado Costa et al. 2018), the reconfirmation of the results permitted evaluation of effects on intraspecific interactions and therefore, the effects on the antipredator mechanism.

INSERT TABLE 1

### 3.2 Experimental stage: Interaction phase

### 3.2.1 Experiment I: Intra- and inter-specific interactions

During the interaction phase, no mortality was observed because of previous exposure of individuals to the insecticide. At the end of the experiment, and regardless of the type of exposure all prey (*R. fernandezae* tadpoles) was consumed. Considering each group separately (control and treated group), the analysis of all variables showed no significant differences between predator-prey proportions, except for the time to eat all prey (Control group: $F(3, 8) = 37.382$, $p < 0.00001$; Treated group: $F(3, 7) = 19.153$, $p < 0.0005$) (Table 2). Therefore, predators took longer times to consume all prey in those treatments with more available items.

When comparing control versus treated groups, we detected significant differences considering the time to eat the first prey (t = -2.168, df = 22, p = 0.041), and the time to eat the second prey (t = -2.258, df = 21, p = 0.035). In this case, tadpoles from the treated group took longer times to eat prey from the proportion 4*Co*-6*Rf* (switch point). However, no significant differences were observed between groups when considering the time to eat a congener (t = 0.394, df = 9, p = 0.703) and the time without eating after depletion of all prey (t = 0.419, df = 4, p = 0.698). These results showed that chlorpyrifos did not affect the efficiency of the antipredator mechanism but did alter the interspecific interactions.

Considering all treatments and recorded data (Table 2), we observed 31 cases of cannibalism, 12 (39%) of which occurred in treated groups and 19 (61%) occurred in control groups. It is important to highlight that all cases of cannibalism occurred after depletion of all prey items, and those cannibalistic events were higher at higher proportions of predators (= lower proportions of prey) as predicted and previously corroborated (Salgado Costa et al. 2014, Salgado Costa 2016). Thus, the prediction was re-confirmed even in the presence of chlorpyrifos. In addition, the differences detected in acoustic variables after the exposure phase (= at 0h of interaction) were maintained during the 24 h of interaction (Np: U = 24.00, p < 0.005; Df: U = 8.50, p < 0.0005). Therefore, the anti-predator mechanism continued to be operating and efficient although sounds were altered.

According to previous descriptions of the antipredator mechanism (Salgado Costa et al. 2014, Salgado Costa 2016), different possible cues could be part of it, among them a rapid movement of the tail and the emission of sounds, added to chemical and visual recognition cues. However, the antipredator mechanism of the species *L. llanensis* was described as operational without acoustic cues under controlled laboratory conditions and without exposure to contaminants (Salgado Costa et al. 2016). The mechanism was described as less efficient than that of its related species *Ceratoprhys cranwelli*. Under this scenario, it is necessary to reinterpret the function of the acoustic signals in each species. Both *C. ornata* and *C. cranwelli* tadpoles appear to use those cues along with others to avoid predation (Salgado Costa et al. 2014, Salgado Costa 2016), while *L. llanensis* has a less complex antipredator mechanism without acoustic cues. Here we demonstrated that although some cues were affected (tail movements, sound emission), the efficiency of the antipredator mechanism of *C. ornata* remained unaltered. Therefore, cues may be acting in a complementary manner since the mechanism remains operational. Future studies should be conducted to determine the presence and importance of each individual cue.

INSERT TABLE 2

Figure 1 shows the cumulative consumed prey and congeners during the experiment, considering the different predator-prey proportions and groups. Since each proportion has a limited number of prey to consume, the variable was standardized considering the percent of prey consumed in relation to the initial number of available prey. Note that we randomly selected a time within a normal appetite period (132 min), and two times within a high appetite period (1354 min and 2871 min) as previously determined for *C. ornata* larvae. The normal appetite period is the time tadpoles spend to eat when they are not starving; cannibalism does not occur although prey items are not available. The high appetite period begins from the end of the normal one and is characterized by the occurrence of events of cannibalism (Salgado Costa 2016). In those times (normal and high appetite period), we compare the cumulative prey and congener consumption within and between treatments (predator-prey proportions). In both groups (control and treated group) significant differences were observed between the different predator-prey proportions at all the analyzed times. Such differences were visualized during the normal appetite period between the proportion 4*Co*-6*Rf* and 8*Co*-2*Rf* (H = 8.154, p = 0.001) in the control group, and between 2*Co*-8*Rf* and 6*Co*-4*Rf* in the treated group (H = 8.752, p = 0.001). During the high appetite period, differences were visualized between the higher proportion of predators (= lower proportion of prey, 8*Co*-2*Rf*) and the higher proportion of prey (2*Co*-8*Rf*) (1354h-control group: H = 10.000, p < 0.0001; 1354h-treated group: H = 9.690, p < 0.0001; 2871h-control group: H = 10.000, p < 0.0001). Though, during the high appetite period, almost 100% of the available prey had been consumed in all the predator-prey proportions. Conversely, during the normal appetite period, all prey had been eaten only in the proportions where less prey was available (8*Co*-2*Rf*, 6 *Co*-4*Rf*; Figure 1).

INSERT FIG 1

Comparison of the normal appetite period (132min) regardless of the *Co-Rf* ratio, indicated that tadpoles exposed to chlorpyrifos consumed significantly less prey (= delay consumption) than the control group (V = 21.00; p = 0.028). These differences were more evident between treatments with more available prey (4*Co*-6*Rf* and 2*Co*-8*Rf*) (Figure 1-A,B).

No significant differences in cannibalistic events were observed during the normal appetite period among the predator-prey proportions of the treated group (H = 3.667, p = 0.160). However, the control group showed significantly different cannibalistic events in the proportion with no prey (10*Co*-0*Rf*) (H = 6.00; p = 0.028) (Figure 1-C,D). Treated and control groups showed significant differences in the percent of eaten congeners (1354min-control: H = 7.146, p = 0.014; 1354min-treated: H = 9.279, p = 0.004; 2871min-control H = 9.467, p = 0.002; 2871min-treated: H = 9.229, p = 0.002). These differences were between higher proportions of predators (= lower proportions of prey). However, the comparison of the number of cannibalistic events for each of these times, regardless of the proportion in which they occurred, did not show significant differences between the exposed larvae and those of the control group (132 min: V = 1.50, p = 1.000; 1354 min: V = 23.00, p = 0.477; 2871 min: V = 15.00, p = 0.343). This result confirms that the insecticide did not affect the antipredator mechanism at any of the analyzed times.

Cannibalistic events are well-known processes that involve both killing and consuming an individual of the same species, this being able to influence life-history, population structure, behavior, and competition for resources. They have been reported in a large number of species, from protozoa to mammals, and occurs frequently in response to low food availability, increased stress and/or high population density (Crump, 1992; Claessen, De Roos and Persson, 2004). The study species has an antipredator mechanism that diminishes the frequency of predation between conspecifics in the presence of heterospecific prey. Chlorpyrifos does not seem to affect the number of cannibalistic events but leads to less prey consumption in the exposed larvae, which could affect their growth, competitive capacity, survival and potentially population dynamics. This could give exposed larvae a competitive disadvantage with other organisms that feed on the same resources (overlapping niches) and are less sensitive or not affected by chlorpyrifos.

### 3.2.2 Experiment 2: Effects on intraspecific interactions

During the 7 h 30 min of visual observations and sound monitoring, 64 intraspecific interactions were recorded (Table 3). The analysis of the contingency table indicated that when individuals in the role of predators belong to the control group (C) there is a relation between the emission/non-emission of sounds and the exposure condition of the individual (control / treated group) ($X^2$ = 9.802, p < 0.05). That is, interactions between control and treated individuals (C-t) were less frequent (= 33%) than those between control individuals (C-c, 67%). In addition, when the interaction was between control individuals, the conspecific prey emitted sounds in 95% of the interactions (n = 37). In contrast, when the interaction was between control and treated individuals (C-t), the conspecific prey of the treated group emitted sounds in only 24% of the interactions (n = 12). This can be attributed to the lower activity of treated individuals (= lower number of interactions between control and treated individuals). Therefore, chlorpyrifos affected intraspecific interactions of *C. ornata* larvae. However, if individuals in the role of predators belong to the treated group (T-c/T-t) there were no significant differences between observed and expected frequencies ($X^2$ = 0.24, p > 0.05). Again, this can be attributed to the lower activity of treated individuals.

INSERT TABLE 3

### 3.3 Theoretical stage: computational modeling

In all cases, we determined the mean number of consumed prey and predators as a function of time (in MC units) over 100 independent experiments. Note that we repeated experiments to assure reliability and independence of the initial (random) conditions. We compared the results of the number of consumed prey from model I (without antipredator mechanism) and model II (presence of an antipredator mechanism) with experimental data (Figure 2-A). We re-scaled the simulation-time-scale to allow the comparison by matching the full-time range of measurements in both simulations and experiments, that is, $t_{allsim}=t_{allexp}$. As a first observation, Model II recapitulated experimental data more accurately (t = -0.561, p = 0.579) than Model I (t = 3.704, p = 0.001) (see also Supplementary Figure 2). We performed the same comparison for the time to eat all prey as a function of the initial number of predators ($N_{co}$, Figure 2-B). Again, Model II recapitulated more accurately experimental data, indicating a fast decrease of all prey with $N_{co}$ (Supplementary Figure 2). Based on these results, it can be concluded that both prey and predators move randomly. Therefore, interactions may be considered as simple encounter issues, where predators eat prey with a high probability and cannibalism only occur when no prey are available, at least in the proximity of the predator.

The simulation allowed us to determine the total number of predator-predator encounters for each independent sample and to compare those with the number of cannibalistic events. The results of the simulation indicated that the number of intraspecific encounters significantly exceeds the number of cannibalisms (t = 2.812, df = 6, p = 0.031; Table 4). The finding was quite surprising as it is in line with published results that indicated that cannibalism does not occur every time two individuals of *Co* interact; rather, other visual, chemical and/or motion signals may be involved (Salgado Costa, 2016). So, we introduced to Model II the possibility to stay in a site instead of moving (Model III = Model II + insecticide). As previously stated, under this new model, a randomly selected tadpole can either jump to an empty neighbor or remain in a site with 1/5 probability. Thus, movements were slowed down. Again, the results of the simulation indicated that the number of intraspecific encounters significantly exceeds the number of cannibalisms (U = 1.00, p = 0.043; Table 4). Therefore, the functionality of the antipredator mechanism under different situations is demonstrated. In addition, figure 3 shows the results of the comparison of the number of consumed prey as a function of time under Model II and III with experimental data (control and treated groups). Experimental data and computational modeling matched (treated group versus Model III: t = 1.487, p = 0.144), especially for $N_{Co} > 2$ (the collapse of computational and experimental data is accurate). Thus, the main effect of chlorpyrifos on predator-prey interactions seems to be the decrease in tadpole movement.

Finally, the simulation allowed us to determine the total number of predator-predator encounters under model III, reproducing the conditions of Experiment II where 3 *Co*-like particles are randomly placed in a LxL=100x100 box (Table 3). The analysis of the contingency table showed no significant differences between the number of encounters registered during the experimental stage and the simulation ($X^2$ = 0.182, p > 0.05). That is, computational modeling reproduced the experimental conditions and results matched within errors.

Therefore, once an adequate model is known, the effects of the alteration of the variable of interest (in our case *C. ornata* predator-prey interactions) due to a stressor (in our case chlorpyrifos) can be predicted without the need to experimentally expose more individuals. Modeling allows obtaining biological data on a large scale that would otherwise be difficult to record (e.g., number of encounters between many predators and prey) maybe because the species is not easy to maintain or breed in controlled conditions, or their conservation status is 'endangered', or simply because a large amount of space and inputs are required, among other reasons. Although it is important to know the study system adequately, once the model is found many modifications can be incorporated (in our case e.g., predator and/or prey mortality, competition, etc.). The aim is to replicate as closely as possible the real conditions under the paradigm of a more emphatic experimentation. In brief, modeling is a new tool that could be used not only to evaluate potential effects of compounds with similar modes of action to CPF on *C. ornata*, but also to model the effects of CPF and related compounds on related species with similar behavior. In this way, the tool would allow preventive conservation decisions to be made even before bioassays are conducted.

INSERT FIGURE 2 AND 3
INSERT TABLE 4

## 4. Conclusions

Chlorpyrifos is a widely used insecticide with a high persistence in water and sediment that reaches aquatic ecosystems. Therefore, the exposure of tadpoles of *C. ornata* is a possible scenario that in case of occurrence could adversely alter their fitness.

Historically, the toxicity of a pollutant (e.g., pesticides) has been studied mainly considering lethal effects. However, the new trends in ecotoxicological studies are giving more emphasis to sublethal effects since they can strongly modify species fitness (Newman, 2014), even more if they occur at higher ecological levels. The present research reports sublethal effects that may compromise the whole population since they occurred at an environmentally relevant concentration.

We found that sublethal environmentally relevant concentrations of chlorpyrifos have the potential to alter intra- and inter-specific interactions of tadpoles, however, they did not affect the antipredator mechanism itself. Intraspecific interactions are the main drivers of population structure and dynamics. They become more relevant when resources are scarce, and competition is greater since they affect the probability of survival and the reproduction of the less fit organisms. The ultimate effect of intraspecific competition is then to decrease the contribution of individuals to the next generation, but also to increase the effectiveness of strong competitors, that is, their proportional contribution to the next generation. Although the experimental data has been generated under laboratory conditions, its comparison with computational models indicates that the effects of chlorpyrifos on intra and interspecific interactions should be considered during risk assessments.

The main effect of chlorpyrifos on predator-prey interactions seems to be the decrease in tadpole movement. Considering that behavioral responses may play an important role in population dynamics and survival, this insecticide may also alter the composition of ecological communities.

In summary, experimental data and computational modeling match accurately, allowing to predict reliable results. Although the Random Walk Model used is the simplest that could have been selected, it accurately and adequately simulated the experimental results. More variables could have been added to this model, however, the more parsimonious option still allows describing more complex and therefore realistic systems. In this sense, computational simulations allow predicting natural processes as an alternative to complex large-scale systems and unethical studies that require to experiment with many living organisms. Here, we provide a valuable ecotoxicological tool of direct practical relevance that is in accordance with the care and welfare of laboratory animals.

**Acknowledgements**

This research was funded by PICT 2015-3137 and PICT 2016-1556 from ANPCyT, and the projects X865 and PPID 80120190500114LP from National University of La Plata. MJPS acknowledges Consejo Interuniversitario Nacional (CIN) for a scholarship for scientific vocations. The support of the National Council of Scientific and Technical Research (CONICET), the National University of La Plata and the University of Ottawa Research Chair in Neuroendocrinology is also acknowledged. Simulations were done on the cluster of Unidad de Cálculo, Instituto de Física de Líquidos y Sistemas Biológicos (IFLYSIB).

Table 1. Mean ± standard deviation of different evaluated endpoints on tadpoles of *Ceratophrys ornata* belonging to a control group (0 mg chlorpyrifos/L) and a treated group (0.02 mg chlorpyrifos/L) considering 96 h of exposure.

| Evaluated Endpoints | Control group | Treated group |
|---|---|---|
| **Mortality (%)** | 10.4 ± 3.8 | 30.6 ± 8.9** |
| **Swimming alterations (%)** | 0 ± 0 | 100 ± 0 |
| **Tail flexure (%)** | 0 ± 0 | 4.3 ± 7.8 |
| **Subcutaneous air (%)** | 0 ± 0 | 100 ± 0 |
| **Prey consumption (%)** | 100 ± 0 | 46.9 ± 27.1** |
| **Sound duration (s)** | 0.028 ± 0.008 | 0.026 ± 0.009 |
| **Number of pulses** | 1.976 ± 0.253 | 1.412 ± 0.515** |
| **Dominant frequency (Hz)** | 4810 ± 1226 | 7977 ± 774** |

**highly significant differences (p < 0.001) between groups.

Table 2. Mean ± standard deviation of each variable under each treatment (= predator-prey proportion). All measures of time are in minutes.

| Variables (min) | Group | Predator-prey proportions | | | | | |
|---|---|---|---|---|---|---|---|
| | | 0Co-10Rf | 2Co-8Rf | 4Co-6Rf | 6Co-4Rf | 8Co-2Rf | 10Co-0Rf |
| 1st p | C | - | 20 ± 3 | 18 ± 1 | 18 ± 1 | 22 ± 0 | - |
| | T | - | 72 ± 46 | 35 ± 11 | 27 ± 11 | 20 ± 8 | - |
| 2nd p | C | - | 29 ± 6 | 25 ± 2 | 25 ± 2 | 32 ± 0 | - |
| | T | - | 162 ± 175 | 86 ± 67 | 42 ± 12 | 55 ± 46 | - |
| All p | C | - | 685 ± 138 | 134 ± 109 | 42 ± 9 | 32 ± 0 | - |
| | T | - | 1499 ± 452 | 638 ± 300 | 65 ± 37 | 55 ± 46 | - |
| We | C | - | - | 989 ± 224 | 627 ± 0 | 1073 ± 1163 | - |
| | T | - | - | - | 2453 ± 0 | 722 ± 865 | - |
| Cong | C | - | - | 1124 ± 136 | 674 ± 0 | 1105 ± 1163 | 494 ± 601 |
| | T | - | - | - | 2561 ± 0 | 777 ± 911 | 350 ± 64 |
| Cnb | C | 0 | 0 | 5 (3/3) | 1 (1/3) | 6 (3/3) | 7 (3/3) |
| | T | 0 | 0 | 0 | 1 (1/3) | 7 (3/3) | 4 (2/3) |

'1st p' = time to eat the first prey; '2nd p' = time to eat the second prey; 'All p' = time to eat all prey; 'We' = time without eating (after depletion of all prey); 'Cong' = time to eat a congener; 'Cnb' = number of events of cannibalism, the number of replicates out of total replicates in which those events occurred are in parenthesis; *Co* = *Ceratophrys ornata* tadpoles in the role of predators; *Rf* = *Rhinella fernandezae* tadpoles in the role of prey; C = control group (0 mg chlorpyrifos/L); T = treated group = (0.02 mg chlorpyrifos/L).

Table 3. Experimental (A) data of the intraspecific interactions of three *Ceratoprhys ornata* tadpoles under different conditions (exposed or not to chlorpyrifos), considering the emission (SE) or no emission of sounds (NSE). The number of predator-predator encounters (*P-P*) is detailed for experimental (A) and modeled (B, model III) data.

| Interactions | A | | | B |
|---|---|---|---|---|
| | SE | NSE | *P-P* | *P-P* |
| **C-c** | 37 | 2 | 39 | 58 |
| **C-t** | 12 | 7 | 19 | 17 |
| **Total** | 49 | 9 | 58 | 75 |
| **T-c** | 4 | 1 | 5 | 4 |
| **T-t** | 1 | 0 | 1 | 2 |
| **Total** | 5 | 1 | 6 | 6 |

C = tadpoles in the role of predators from the control group; c = tadpoles in the role of prey from the control group; T = tadpoles in the role of predators from the treated group; tadpoles in the role of prey from the treated group.

Table 4. Mean ± standard deviation of the number of predator-predator encounters (*P-P*) and the number of events of cannibalism (*Cnb*) under Model II (with the presence of an antipredator mechanism) and Model III (Model II + insecticide), considering different predator-prey proportions.

|  | Variables | Predator-prey proportions | | | |
|---|---|---|---|---|---|
|  |  | *2Co-8Rf* | *4Co-6Rf* | *6Co-4Rf* | *8Co-2Rf* |
| **Model II** | *P-P* | 5±6 | 16±13 | 24±18 | 28±22 |
|  | *Cnb* | 0.5±0.5 | 2.5±0.7 | 4.6±0.6 | 6.5±0.6 |
| **Model III** | *P-P* | 1±2 | 5±5 | 10±7 | 15±12 |
|  | *Cnb* | 0 | 0 | 0.2±0.5 | 1.8±1.5 |

*Co* = *Ceratophrys ornata* tadpoles in the role of predators; *Rf* = *Rhinella fernandezae* tadpoles in the role of prey

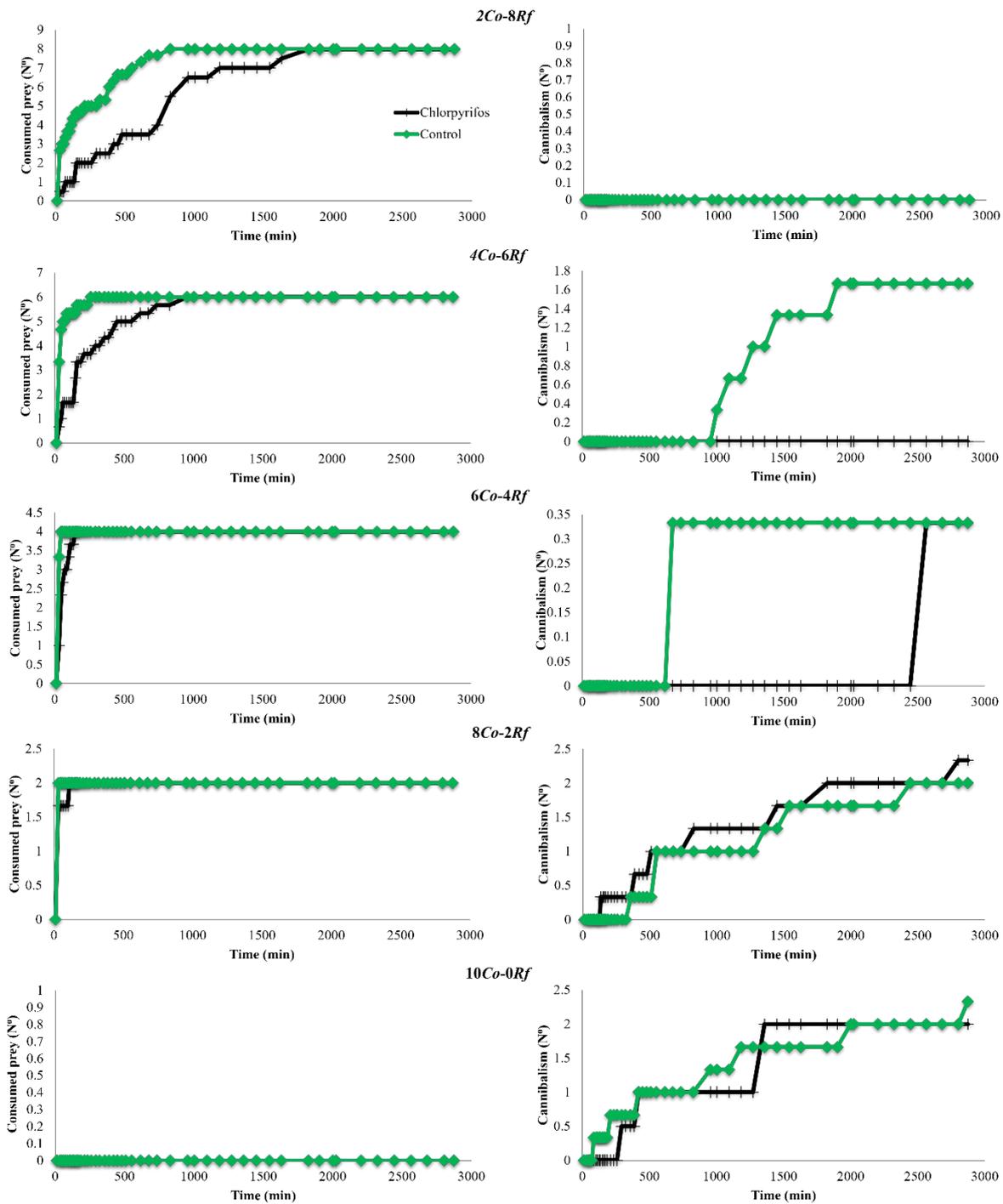

Figure 1. *Ceratoprhys ornata* (*Co*) and *Rhinella fernandezae* (*Rf*, prey) interactions along the time of the experiment, considering different predator-prey proportions. Left panel: accumulated mean of consumed prey (*Rf*) by predators (*Co*) considering predators from the control group (0 mg/L of chlorpyrifos, green line), and the treated group (0.02 mg/L of chlorpyrifos, black line); Right panel: accumulated mean of cannibalism considering individuals of *Co* from the control group and the treated group.

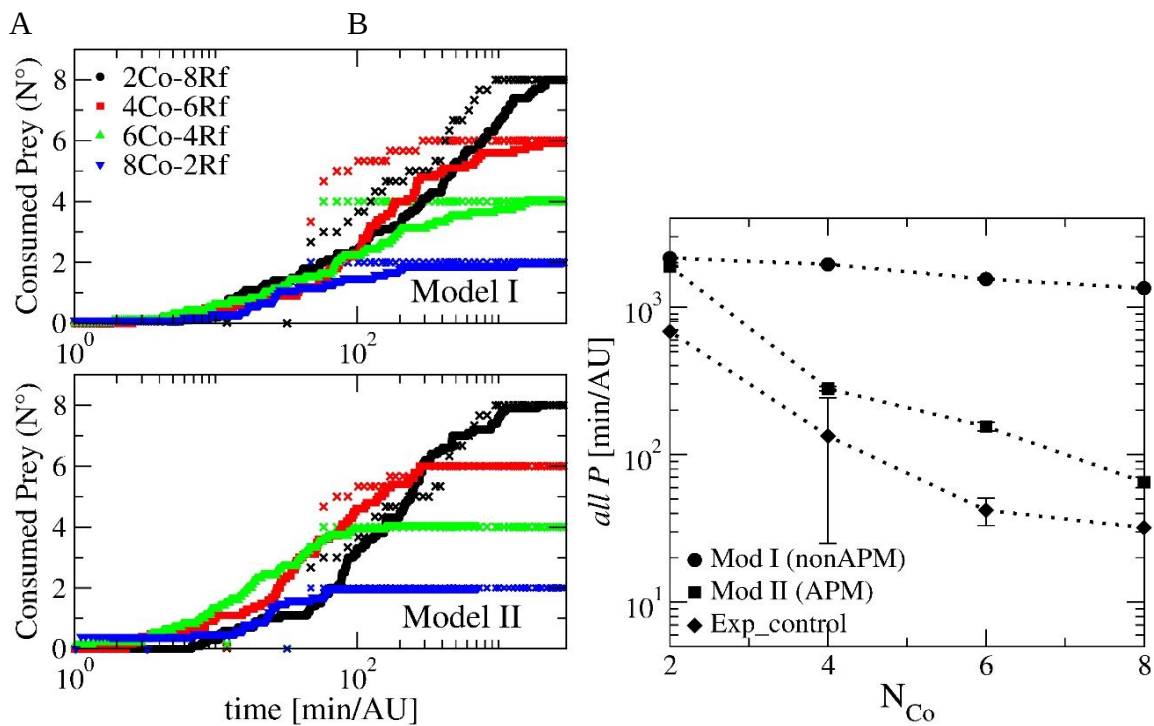

Figure 2. Linear-log plot of A) the number of consumed prey of *Rhinella fernandezae* (*Rf*) under different predator-prey proportions over time (min/AU), and B) the time to eat all prey (*all P*) as a function of the initial number of predators ($N_{Co}$), considering two theoretical models (Mod I and II) and experimental data of the control group (crosses, Exp_Control). Model I is the theoretical model without presence of an antipredator mechanism (nonAPM), Model II is the theoretical model with an antipredator mechanism (APM). Error bars (B) were calculated over the different independent experiments performed (100 simulations, and 10 experiments).

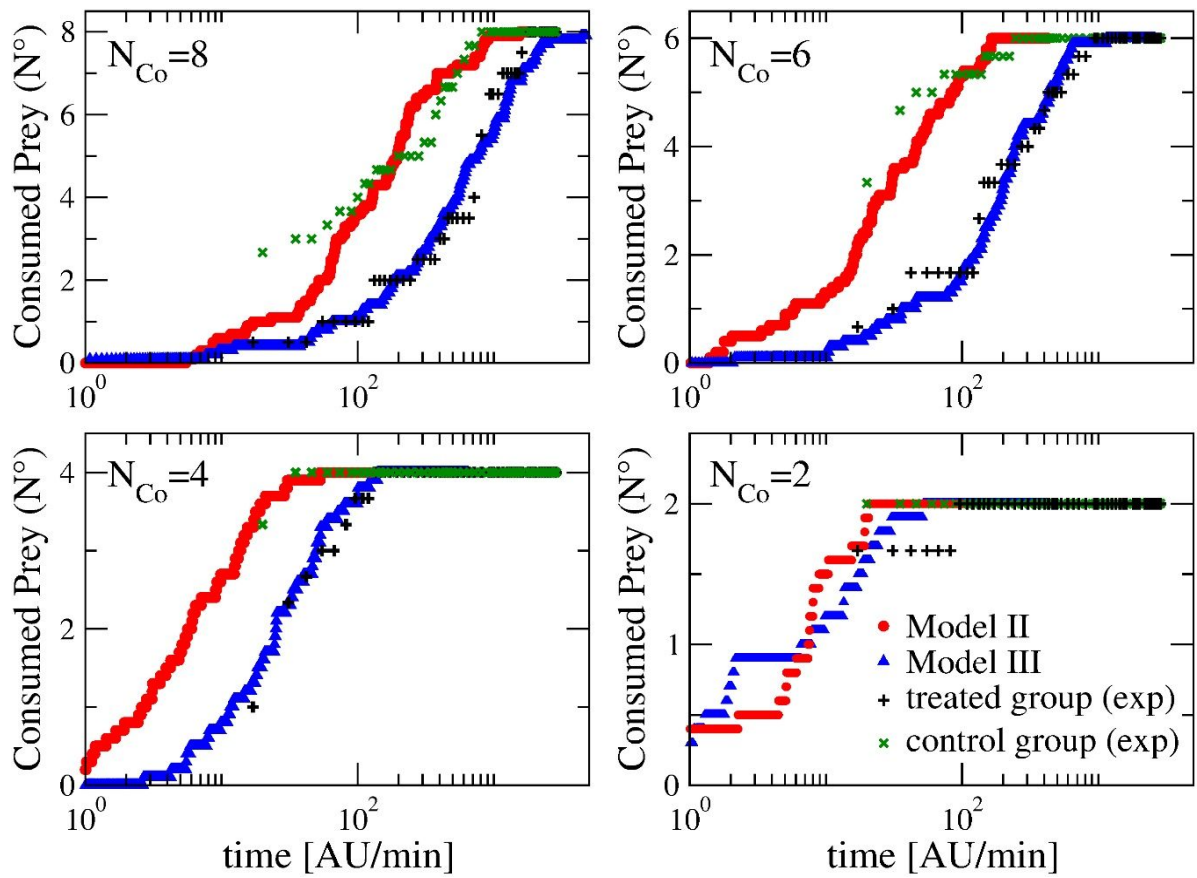

Figure 3. Linear-log plots of consumed prey as a function of time considering different computational models (II and III) and experimental data (black and green crosses).

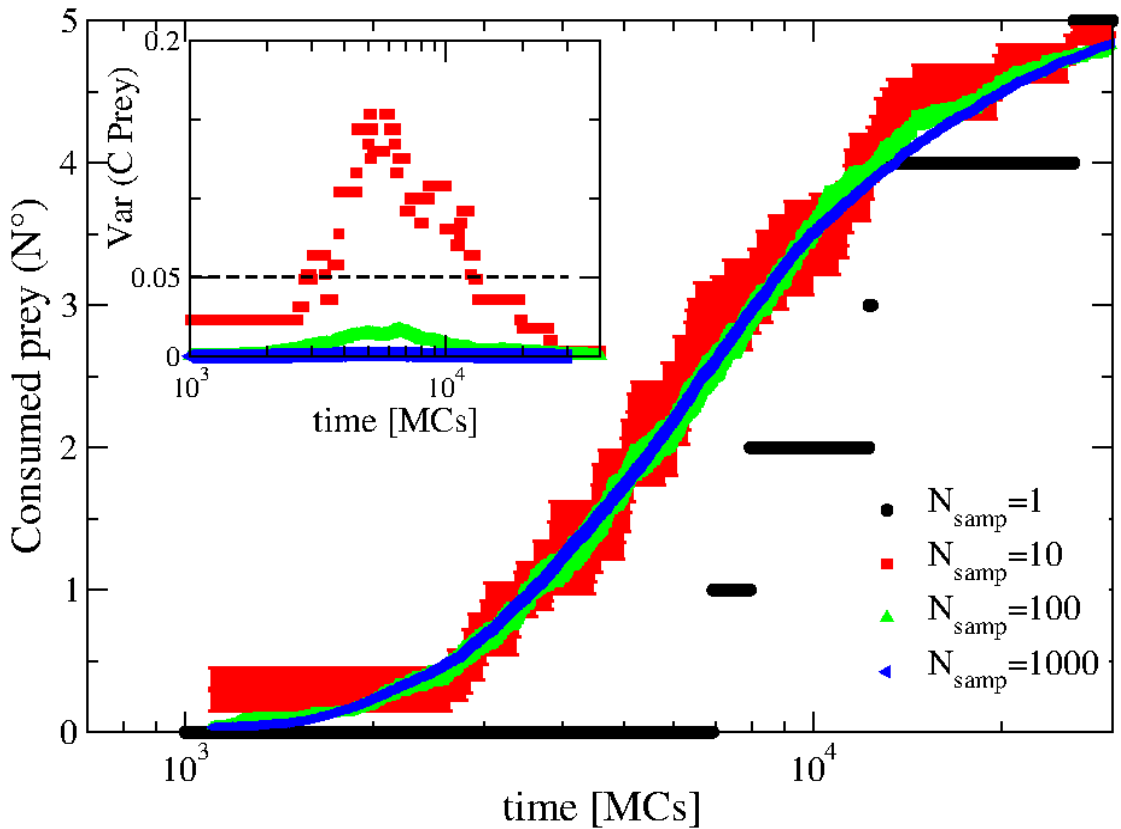

Supplementary Figure 1. Typical results obtained from the number of consumed prey as a function of time in Monte Carlo method (MCs) for $N_{Co}/N_{Rf} = 4/6$, and different independent samples ($N_{samp}$). Error bars are shown. Inset: the variance of the measurement determined for different $N_{samp}$ is less than 5% for all times when $N_{samp} \geq 100$.

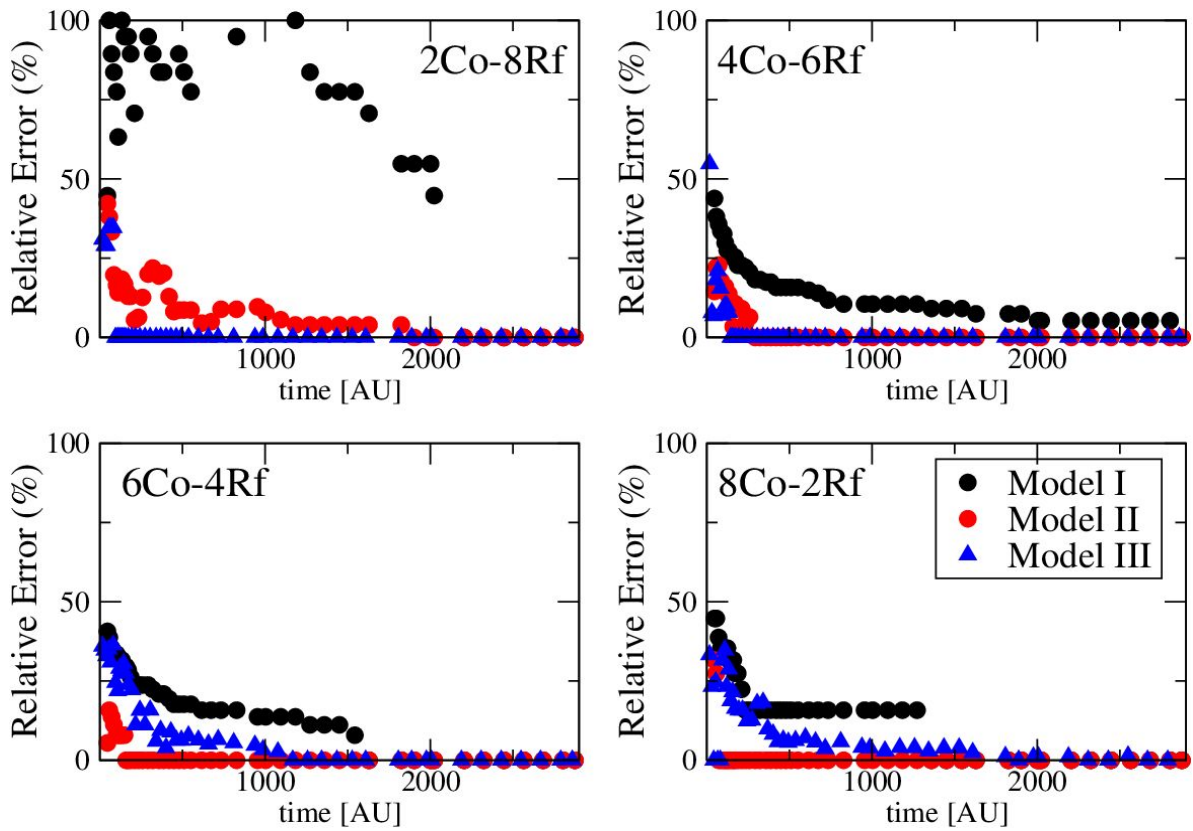

Supplementary Figure 2. Relative percentage error of the number of consumed prey comparing each model (I, II and III) with the experimental data for the different predator-prey proportions as a function of time. *Co = Ceratophrys ornata* tadpoles in the role of predators; *Rf = Rhinella fernandezae* tadpoles in the role of prey.

# References

Alvarez, Melina, Cecile Du Mortier, Soledad Jaureguiberry, and Andrés Venturino. 2019. "Joint probabilistic analysis of risk for aquatic species and exceedence frequency for the agricultural use of chlorpyrifos in the Pampean region, Argentina." *Environmental Toxicology and Chemistry*.

ASTM. (2002). *Standard guide for collection, storage, characterization, and manipulation of sediments for toxicological testing*.

Attademo, Andrés Maximiliano, Paola Mariela Peltzer, Rafael Carlos Lajmanovich, Mariana Cabagna-Zenklusen, Celina María Junges, Eduardo Lorenzatti, Carolina Aró, and Paula Grenón. 2015. "Biochemical changes in certain enzymes of *Lysapsus limellium* (Anura: Hylidae) exposed to chlorpyrifos." *Ecotoxicology and Environmental Safety* 113:287-294.

Bantle, John A, James N Dumont, Robert A Finch, and Gregory Linder. 1996. *Atlas of abnormalities: a guide for the performance of FETAX*. 2nd ed: Oklahoma State University, Stillwater.

Barreto, E., C. Salgado Costa, P. Demetrio, C. Lascano, A. Venturino, and G. S. Natale. 2020. "Sensitivity of *Boana pulchella* (Anura: Hylidae) tadpoles to environmentally relevant concentrations of chlorpyrifos: effects at the individual and biochemical level." *Environmental Toxicology and Chemistry* 39 (4):834-841.

Barron, Mace G, and Kent B Woodburn. 1995. "Ecotoxicology of chlorpyrifos." In *Reviews of Environmental Contamination and Toxicology*, 93. Springer.

Bernabò, Ilaria, Emilio Sperone, Sandro Tripepi, and Elvira Brunelli. 2011. "Toxicity of chlorpyrifos to larval *Rana dalmatina*: acute and chronic effects on survival, development, growth and gill apparatus." *Archives of environmental contamination and toxicology* 61 (4):704-718.

Brewer, SK, EE Little, AJ DeLonay, SL Beauvais, SB Jones, and Mark R Ellersieck. 2001. "Behavioral dysfunctions correlate to altered physiology in rainbow trout (*Oncorynchus mykiss*) exposed to cholinesterase-inhibiting chemicals." *Archives of Environmental Contamination and Toxicology* 40 (1):70-76.

Brunelli, Elvira, Ilaria Bernabò, Cecilia Berg, Katrin Lundstedt-Enkel, Antonella Bonacci, and Sandro Tripepi. 2009. "Environmentally relevant concentrations of endosulfan impair development, metamorphosis and behaviour in *Bufo bufo* tadpoles." *Aquatic Toxicology* 91 (2):135-142.

Cei, José M. 1980. *Amphibians of Argentina*. Edited by Università degli Studi di Firenze. Vol. Monogr. 2. Firenze, Italia: Monitore Zoologico Italiano.

Claver, Ana, Peña Ormad, Luis Rodríguez, and José Luis Ovelleiro. 2006. "Study of the presence of pesticides in surface waters in the Ebro river basin (Spain)." *Chemosphere* 64 (9):1437-1443.

Cowman, DF, and LE Mazanti. 2000. "Ecotoxicology of ''new generation'' pesticides to amphibians." *Ecotoxicology of amphibians and reptiles*:233-268.

DeGarady, Colette J, and Richard S Halbrook. 2006. "Using anurans as bioindicators of PCB contaminated streams." *Journal of Herpetology* 40 (1):127-130.

El-Merhibi, A, Anupama Kumar, and T Smeaton. 2004. "Role of piperonyl butoxide in the toxicity of chlorpyrifos to *Ceriodaphnia dubia* and *Xenopus laevis*." *Ecotoxicology and environmental safety* 57 (2):202-212.

Gallardo, José María. 1974. *Anfibios de los alrededores de Buenos Aires*: Editorial Universitaria de Buenos Aires.

Gosner, Kenneth L. 1960. "A simplified table for staging anuran embryos and larvae with notes on identification." *Herpetologica* 16 (3):183-190.

Jergentz, S, H Mugni, C Bonetto, and R Schulz. 2005. "Assessment of insecticide contamination in runoff and stream water of small agricultural streams in the main soybean area of Argentina." *Chemosphere* 61 (6):817-826.

Kavitha, P, and J Venkateswara Rao. 2008. "Toxic effects of chlorpyrifos on antioxidant enzymes and target enzyme acetylcholinesterase interaction in mosquito fish, *Gambusia affinis*." *Environmental Toxicology and Pharmacology* 26 (2):192-198.

Kerby, Jacob Lawrence. 2006. "Pesticide effects on amphibians: A community ecology perspective." University of California, Davis.

Landau, David P, and Kurt Binder. 2014. *A guide to Monte Carlo simulations in statistical physics*. Cambridge: Cambridge University Press.

Liendro, Natacha, Ana Ferraria, Mariana Mardirosian, Cecilia I. Lascano, and Andrés Venturino. 2015. "Toxicity of the insecticide chlorpyrifos to the South American toad *Rhinella arenarum* at larval developmental stage." *Environmental Toxicology and Pharmacology* 39 (2):525-535.

Mac Loughlin, Tomás M., Leticia Peluso, and Damián J. G. Marino. 2017. "Pesticide impact study in the peri-urban horticultural area of Gran La Plata, Argentina." *Science of the Total Environment* 598:572-580.

Marino, D, and A Ronco. 2005. "Cypermethrin and chlorpyrifos concentration levels in surface water bodies of the Pampa Ondulada, Argentina." *Bulletin of Environmental Contamination and Toxicology* 75 (4):820-826.

Natale, G.S., and C. Salgado Costa. 2012. "*Ceratophrys ornata* (Bell, 1843). Escuerzo común." In *Categorización del estado de conservación de los anfibios de la República Argentina*, 131-159. Cuadernos de herpetología.

Natale, Guillermo S, Leandro Alcalde, Raul Herrera, Rodrigo Cajade, Eduardo F Schaefer, Federico Marangoni, and Vance L Trudeau. 2011. "Underwater acoustic communication in the macrophagic carnivorous larvae of *Ceratophrys ornata* (Anura: Ceratophryidae)." *Acta Zoologica* 92 (1):46-53.

Newman, Michael C. 2014. *Fundamentals of ecotoxicology: the science of pollution*. Boca Raton, Florida: CRC press.

Quiroga, Lorena B, Eduardo A Sanabria, Miguel W Fornés, Daniel A Bustos, and Miguel Tejedo. 2019. "Sublethal concentrations of chlorpyrifos induce changes in the thermal sensitivity and tolerance of anuran tadpoles in the toad *Rhinella arenarum*?" *Chemosphere* 219:671-677.

Relyea, Rick A, and Kerry Edwards. 2010. "What doesn't kill you makes you sluggish: how sublethal pesticides alter predator–prey interactions." *Copeia* 2010 (4):558-567.

Richards, Sean M, and Ron J Kendall. 2002. "Biochemical effects of chlorpyrifos on two developmental stages of *Xenopus laevis*." *Environmental Toxicology and Chemistry* 21 (9):1826-1835.

Ruiz de Arcaute, C., C. Salgado Costa, P. M. Demetrio, G. S. Natale, and A. E. Ronco. 2012. "Influence of existing site contamination on sensitivity of *Rhinella fernandezae* (Anura, Bufonidae) tadpoles to Lorsban®48E formulation of chlorpyrifos." *Ecotoxicology* 21 (8):2338-48.

Salgado Costa, C., A. E. Ronco, V. L. Trudeau, D. Marino, and G. S. Natale. 2018. "Tadpoles of the horned frog *Ceratophrys ornata* exhibit high sensitivity to chlorpyrifos for conventional ecotoxicological and novel bioacoustic variables." *Environmental Pollution* 235:938-947.

Salgado Costa, Carolina. 2016. "Desarrollo de un modelo experimental con larvas de *Ceratophrys* spp. (Anura: Ceratophryidae) para su aplicación en estudios ecotoxicológicos de plaguicidas: efectos sobre variables convencionales y bioacústicas." PhD, Facultad de Ciencias Naturales y Museo. Universidad Nacional de La Plata.

Salgado Costa, Carolina, Mariana Chuliver Pereyra, Leandro Alcalde, Raúl Herrera, Vance L. Trudeau, and Guillermo S. Natale. 2014. "Underwater sound emission as part of an antipredator mechanism in *Ceratophrys cranwelli* tadpoles." *Acta Zoologica* 95 (3):367-374. doi: 10.1111/azo.12035.

Salgado Costa, Carolina, Vance Lionel Trudeau, Alicia Estela Ronco, and Guillermo Sebastián Natale. 2016. "Exploring antipredator mechanisms: New findings in ceratophryid tadpoles." *Journal of Herpetology* 50 (2):233-238. doi: 10.1670/14-179.

Sansiñena, Jesica A., Leticia Peluso (Eq), Carolina Salgado Costa (Eq), Pablo M. Demetrio, Tomás M. Mac Loughlin, Damián J.G. Marino, Leandro Alcalde, and Guillermo S. Natale. 2018. "Evaluation of the toxicity of the sediments from an agroecosystem to two native species, *Hyalella curvispina* (CRUSTACEA: AMPHIPODA) and *Boana pulchella (*AMPHIBIA: ANURA), as potential environmental indicators." *Ecological Indicators* 93:100-110.

Silva, Márcio Borba, Ricardo Evangelista Fraga, Flávia Lopes Silva, Luana Alzira Alves Oliveira, Tiago Sousa De Queiroz, Mariane Amorim Rocha, and Flora Acuña Juncá. 2020. "Effects of acute exposure of chlorpyrifos on the survival, morphology and swimming ability of Odontophrynus carvalhoi tadpoles." *Ecotoxicology and Environmental Contamination* 15 (1):37-42.

Sotomayor, Verónica, Tai S Chiriotto, Ana M Pechen, and Andrés Venturino. 2015. "Biochemical biomarkers of sublethal effects in *Rhinella arenarum* late gastrula exposed to the organophosphate chlorpyrifos." *Pesticide Biochemistry and Physiology* 119:48-53.

Sotomayor, Verónica, Cecilia Lascano, Ana María Pechen de D'Angelo, and Andrés Venturino. 2012. "Developmental and polyamine metabolism alterations in *Rhinella arenarum* embryos exposed to the organophosphate chlorpyrifos." *Environmental Toxicology and Chemistry* 31 (9):2052-2058.

Testai, Emanuela, Franca M Buratti, and Emma Di Consiglio. 2010. "Chlorpyrifos." In *Hayes' Handbook of Pesticide Toxicology*, 1505-1526. Elsevier.

Trudeau, Vance L., Gustavo M. Somoza, Guillermo S. Natale, Bruce Pauli, Jacqui Wignall, Paula Jackman, Ken Doe, and Fredrick W. Schueler. 2010. "Hormonal induction of spawning in 4 species of frogs by coinjection with a gonadotropin-releasing hormone agonist and a dopamine antagonist." *Reproductive Biology and Endocrinology* 8 (1):36.

USEPA. (2002). *Methods for measuring the acute toxicity of effluents and receiving waters to freshwater and marine organisms*.

Winsberg, Eric. 2019. *Computer Simulations in Science*. Winter 2019 Edition ed: The Stanford Encyclopedia of Philosophy.

Yin, XiaoHui, GuoNian Zhu, Xian Bing Li, and ShaoYing Liu. 2009. "Genotoxicity evaluation of chlorpyrifos to amphibian Chinese toad (Amphibian: Anura) by comet assay and micronucleus test." *Mutation Research/Genetic Toxicology and Environmental Mutagenesis* 680 (1):2-6.